# Self-modulated laser wakefield accelerators as x-ray sources

N Lemos[1], J L Martins[3], F S Tsung[2], J L Shaw[1], K A Marsh[1], F Albert[4], B B Pollock[4] and C Joshi[1]

[1]Department of Electrical Engineering, University of California Los Angeles, 405 Hilgard Ave, Los Angeles CA 90095, USA
[2]Department of Physics and Astronomy, University of California Los Angeles, 405 Hilgard Ave, Los Angeles CA 90095, USA
[3]GoLP/Instituto de Plasmas e Fusao Nuclear, Instituto Superior Tecnico (IST), Universidade Tecnica de Lisboa (UTL), 1049-001 Lisbon, Portugal
[4]Lawrence Livermore National Laboratory, 7000 East Avenue, Livermore CA 94550, USA

Email: nuno.lemos@ucla.edu

**Abstract.** The development of a directional, small-divergence, and short-duration picosecond x-ray probe beam with an energy greater than 50 keV is desirable for high energy density science experiments. We therefore explore through particle-in-cell (PIC) computer simulations the possibility of using x-rays radiated by betatron-like motion of electrons from a self-modulated laser wakefield accelerator as a possible candidate to meet this need. Two OSIRIS 2D PIC simulations with mobile ions are presented, one with a normalized vector potential $a_0 = 1.5$ and the other with an $a_0 = 3$. We find that in both cases direct laser acceleration (DLA) is an important additional acceleration mechanism in addition to the longitudinal electric field of the plasma wave. Together these mechanisms produce electrons with a continuous energy spectrum with a maximum energy of 300 MeV for $a_0 = 3$ case and 180 MeV in the $a_0 = 1.5$ case. Forward-directed x-ray radiation with a photon energy up to 100 keV was calculated for the $a_0 = 3$ case and up to 12 keV for the $a_0 = 1.5$ case. The x-ray spectrum can be fitted with a sum of two synchrotron spectra with critical photon energy of 13 and 45 keV for the $a_0$ of 3 and critical photon energy of 0.3 and 1.4 keV for $a_0$ of 1.5 in the plane of polarization of the laser. The full width at half maximum divergence angle of the x-rays was 62 x 1.9 mrad for $a_0 = 3$ and 77 x 3.8 mrad for $a_0 = 1.5$.

## 1. Introduction

Understanding material properties under extreme conditions of temperature, pressure, and density is essential for different fields of physics such as astrophysics, high energy density science (HEDS), and inertial confinement fusion (ICF). The development of directional, low-divergence, and short-duration (ps and sub-ps) x-ray probes with energies larger than 50 keV is desirable for these applications [1, 2]. For instance, such a high-energy x-ray probe beam could be used to radiograph the imploding inertially-confined fusion capsule at the National Ignition Facility [3, 4] or warm-dense matter created using lasers or Z-pinches via absorption spectroscopy [5] or scattering techniques [6, 7].

Many of the facilities where HEDS is studied either have or are building ps-pulse duration, kJ-class lasers that can be overlapped in space and synchronized time to the typically tens-of-nanosecond-long drive pulse. Among the ideas being explored for producing the x-rays are K-alpha sources [8] and betatron radiation sources.

Until now, betatron x-ray radiation has been mainly produced using either ultra-relativistic electron beams as a driver [9] or using sub-100-fs-duration titanium-sapphire lasers [10-15]. We are proposing to extend the development of the betatron x-ray source using the so-called self-modulated laser wakefield acceleration (SM-LWFA) regime with longer laser pulses to support its applications on large-scale facilities such as Omega-EP, Petal, and NIF-ARC. Due to its directionality, relatively small divergence angle, and high-brightness, an x-ray source based on the oscillation motion of relativistic electrons in a plasma may be an ideal probe and backlighter for time-resolved spectroscopy, imaging, Compton radiography, and Laue diffraction. Some work has been done to understand such "betatron-like" x-ray sources produced with 0.5-1 ps laser pulses [16]. In just one previous experiment on the Vulcan facility, where a 0.7 ps pulse was focused to give a normalized vector potential $a_0 = eA/mc^2$ of between 9 and 29, the laser pulse produced a plasma channel and x-rays were thought to be generated by the process of direct laser acceleration (DLA) [16]. Although the ps class laser facilities mentioned above will contain a great deal of energy, the peak $a_0$ will be typically less than 5 due to the long focusing optics. In this paper, we therefore intentionally propose to use much lower $a_0$ laser pulses to investigate the regime where the laser pulse can undergo self-modulation [17-20] and the forward Raman scattering instability [21, 22] before channel formation can occur.

The goal is to explore whether the total accelerated charge and thereby the x-ray photon yield can be increased by self-trapping the plasma electrons in tens of plasma wavelengths instead of just one as in LWFA with a fs-laser pulse. Furthermore, since the laser pulse will naturally overlap many plasma wavelengths and therefore the accelerating electrons, there is expected to be a significant contribution from DLA [23, 24] with the possibility of a further increase in the maximum electron energy beyond the so-called dephasing limit [25]. Our particle-in-cell (PIC) simulations show that for $a_0$ of 1.5, the laser pulse indeed undergoes a self-modulation and/or Raman forward instability that produces a relativistic plasma wave that grows from noise to large amplitudes and traps plasma electrons. These electrons gain energy both from the longitudinal field of the plasma wave and from their interaction with the laser field (DLA) [23, 24, 26] and produce x-rays during the acceleration process due to oscillations in the plasma with a synchrotron-like spectrum reaching energies up to 12 keV. These simulation results are in reasonable agreement with the experimental measurements published elsewhere [27]. For a laser pulse with $a_0$ of 3, SM-LWFA leads to the trapping of electrons; however, DLA is the main mechanism for the energy gain of electrons. The generated x-rays reach energies up to 100 keV

This paper is organized as follows. In Sec. 2, the simulation setup is described. Sec. 3 describes the physical mechanisms of electron acceleration and x-ray emission. In Sec. 4, we analyze the simulation results. The radiation generated by the accelerated electrons is presented in Sec. 5, and in Sec. 6, we state the conclusions.

**2. Simulation setup**

The 2D code OSIRIS 3.0 [28] was used to study the physical mechanisms of laser-plasma instabilities when ps, 100-TW-class laser pulses propagate through underdense plasmas. While such simulations have been carried out before with both fixed and moving ions [29, 30], here we use OSIRIS to first work out the accelerated electron spectrum, then tag representative particles at various energies, and then rerun the simulation to record the position, momentum, and fields sampled by the particles at every time step. We then examine the trajectories of the selected electrons and calculate the radiation generated during the acceleration process. For that calculation, we use a post-processing code named JRad [31] that generates the spectral and spatial distribution of the betatron x-rays.

The OSIRIS simulations have been carried out in the speed-of-light frame (moving window) with a box that was 500 x 150 $\mu$m with a resolution of 60 cells/$\lambda$ in the longitudinal direction and 7.2 cells/$\lambda$ in the transverse direction, where $\lambda$ stands for the laser wavelength.

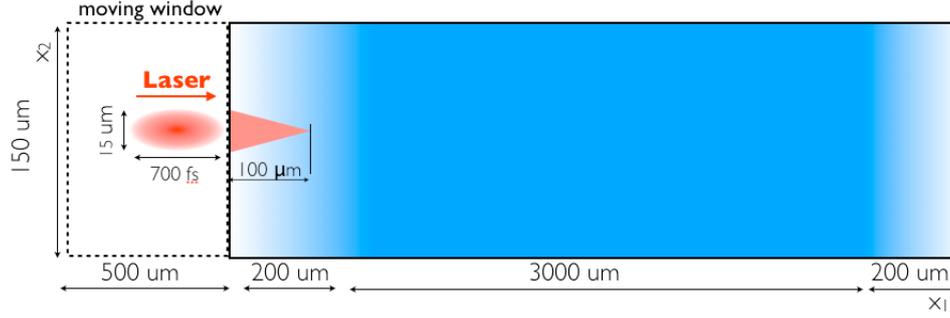

Figure 1 - Layout of the 2D PIC simulation in the speed-of-light frame with a box that was 500 x 150 μm with 30000 x 1024 cells. The laser had an $a_0$ = 1.5 or 3 and was focused 100 μm into the 200-μm-long density up-ramp in a region of fully-ionized He plasma. The density ramp was followed by a 3000 μm constant density region and end with a 200-μm-long-density down ramp.

Figure 1 shows the layout of the 2D simulation. The laser is initialized in vacuum on the right-hand side of the moving window. The moving window containing the laser pulse is launched from the left-hand side into a region of fully-ionized He plasma where the electrons and ions are allowed to move. The plasma density was chosen to be $1.0 \times 10^{19}$ cm$^{-3}$ because this density is approximately where (according to the 1D spatio-temporal theory) the Raman forward scattering (RFS) instability has the highest growth rate for a focused laser $a_0$ of 1-3 [29]. Indeed, previous 1D simulation studies done in Ref. [29] show that the highest-energy electrons were obtained in a range of densities around $1.0 \times 10^{19}$ cm$^{-3}$. The laser pulse was linearly polarized (in plane of the simulation in the $x_2$ direction, where $x_1$ is the direction of the laser propagation) with a pulse duration of 700 fs and a wavelength of 1053 nm. It was focused in the middle of the density up-ramp with a spot size of 15 $\mu$m (FWHM of the electric field) and a peak $a_0$ of 3 or 1.5. In the rest of this paper, we will address the simulation with an $a_0$ of 3 by Sim.1 and with an $a_0$ of 1.5 by Sim.2.

### 3. Physical mechanisms of electron acceleration and x-ray emission

When a high-intensity ($a_0 > 1$) laser pulse with a pulse duration $\tau_p \gg$ plasma period $\lambda_p$ propagates through an underdense plasma ($n_e < n_c$), it can self-focus [32,33] and can also (as stated above) generate plasma waves through the RFS instability [22]. The laser will self focus if $P/P_c > 1$, where P is the laser peak power, $P_c = 17 \, n_c / n_e$ [GW] is the critical power for relativistic self-focusing [32, 33], $n_c$ is the critical density (where the laser frequency $\omega_0$ is equal to the plasma frequency $\omega_p = (4\pi n_e e^2/m_e)^{1/2}$), $n_e$ is the electron plasma density, e is the electron charge, and $m_e$ is the electron mass. The experimental evidence for the self-modulation (SM) and RFS instability is the generation of sideband frequencies, which arise as a result of energy and momentum conservation. In the forward direction, the incident electromagnetic wave at frequency and wave number ($\omega_0$, $k_0$) decays into a co-propagating electrostatic wave with a phase velocity near the speed of light c (i.e. a plasma wave) with a frequency and wave number ($\omega_p$, $k_p$) and two co-propagating electromagnetic waves referred to as the Stokes for ($\omega_0 - n\omega_p$, $k_0 - nk_p$) and the anti-Stokes for ($\omega_0 + n\omega_p$, $nk_p + k_0$). Here, n stands for the harmonic number of the RFS satellites. The plasma wave can be driven to wave breaking limits [34, 35], trap relativistic background electrons, and accelerate them to high energies.

In 2D, there is no simple way to distinguish between the RFS and SM instabilities in the exact forward direction. The stimulated Raman scattering instability in general will develop over a wide range of angles with each angle having its own threshold intensity. Here, we call the RFS occurring over a small range of angles in the forward direction the self-modulation instability. The exact RFS typically has the highest intensity threshold because the plasma waves associated with RFS are relativistic and therefore have the smallest wavenumbers. The overall plasma wave spectrum has not only a longitudinal field but also a transverse focusing and defocusing field. Self-modulation of the laser pulse occurs because the plasma wave produces regions of low and high electron density that act as focusing and defocusing, respectively, regions for the laser beam. At every crest of the plasma wave (high-density region), the laser beam will defocus and in between, it tends to focus (low-electron-density region). It is this periodic focusing and defocusing that produces a transverse modulation of the laser intensity envelope. In the frequency domain even in 1D, the photon acceleration and deceleration in the plasma wave [36] leads to the creation of the Stokes' and anti-Stokes' sidebands.

The longitudinal electric field first accelerates the self-trapped electrons to relativistic energies, but it can eventually decelerate the electrons as they overtake the wave (i.e. dephase). The transverse field leads to betatron-like oscillations of the off-axis electrons, which causes them to radiate photons in the forward direction. The electron trajectories are not exactly sinusoidal about the axis as for idealized betatron motion of an electron. But we will henceforth call the mechanism for radiation emission betatron radiation for simplicity. The trapped electrons that undergo betatron oscillations in the polarization plane of the laser will see an additional electric field from the laser. This transverse electric field of the laser, when in near resonance with the betatron motion of the electrons [37], will in turn increase the transverse momentum of the trapped electrons. The 1D resonance condition is $N\omega_\beta = (1-v_\parallel/v_\Phi)\omega_0$, where $\omega_\beta$ is the betatron frequency, N is a harmonic of the betatron frequency, $v_\parallel$ is the longitudinal velocity of the electrons, $v_\Phi$ is the phase velocity of the laser, and $\omega_0$ is the laser frequency . The increase in transverse momentum can then be converted into longitudinal momentum via the **v** x **B** force. This process is analogous to the inverse free electron laser acceleration [38, 39] or the inverse ion channel laser acceleration [40] mechanisms that facilitate an energy exchange between the laser field and the electron, but in the laser-plasma literature it is referred to as the direct laser acceleration DLA mechanism [23].

While the coupling to the plasma wave and to the laser field determines an electron's energy and trajectory, the betatron motion of the electron principally leads to radiation loss. The dependence of the betatron radiation in an ion column on an electron's energy, its initial position, and the transverse focusing force is the subject of numerous recent publications [10-15] and will not be repeated here. As we will show, electrons experience different acceleration and focusing conditions depending on where they propagate relative to the laser and the longitudinal electric field. Therefore, the only way to accurately calculate the emitted x-ray spectrum is to follow the actual trajectories of the representative electrons and calculate the radiated power per frequency and solid angle interval using the classical theory of electromagnetism [41].

## 4. Simulation results

We begin by analyzing the evolution of the laser and plasma structure to investigate what types of acceleration structures are generated. Then, we determine how the electrons in each of those structures are accelerated. Figures 2a and 2b show the envelope of the laser electric field for Sim.1 and Sim.2 after the laser propagates 1 mm into the plasma. In both simulations, the front of the laser self-focuses (because $P/P_c$ is 23 for Sim.1 and 6 for Sim.2). Figures 2c and 2d show a lineout of the longitudinal electric field $E_1$. In the case of Sim.1, it is clear that in the region between $750 < x_1(\mu m) < 800$, and somewhat later for Sim.2, strong non-linear plasma waves are being driven at the front of the laser pulse and contribute to the self-guiding [32, 33]. The back of the laser pulse is not driving such strong plasma waves in both cases because the near-axis plasma electrons have been blown out by relativistic self-focusing and the ponderomotive force of the plasma wave [42] itself as seen in figures 2e and 2f. Towards the back of the laser pulse, the electrons have already pulled the ions radially outward (figures 2g and 2h), creating a near-hollow waveguide [30] for the rest of the laser pulse (50% drop in electron and ion density). This density depression is shown in figures 2e and 2g. In the case of Sim.2, the physics is very similar, but because the laser has a lower $a_0$, the spatial temporal gain of the RFS [43] is one hundred times higher than in Sim.1, producing linear plasma waves at the back of the laser pulse ($500<x_1(\mu m)<750$) as shown in figure 2d and 2f. These linear plasma waves are now starting to self modulate the laser within the channel at the plasma frequency as shown in figure 2b.

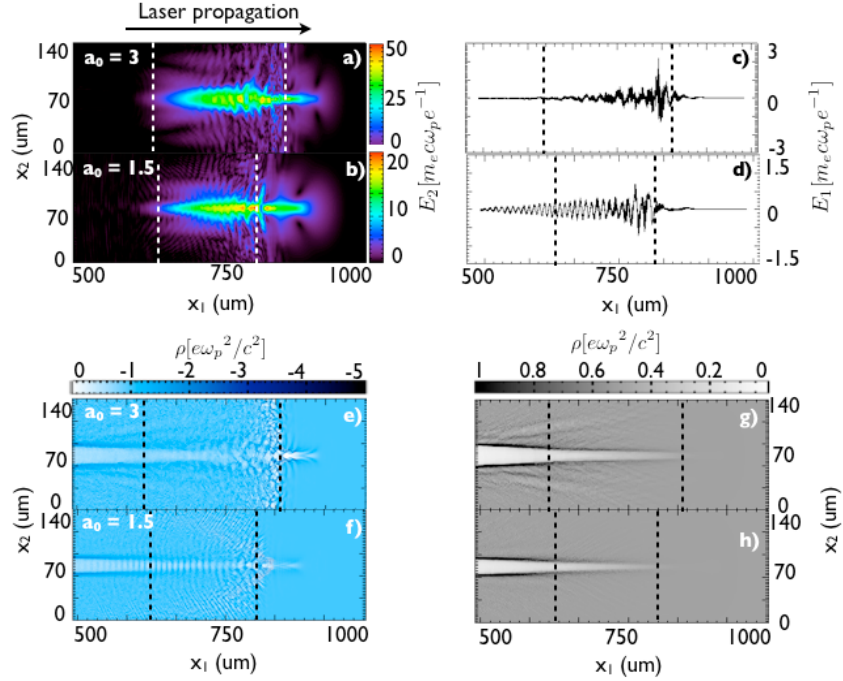

Figure 2 - a) and b) represent the laser electric field envelope for Sim.1 and Sim.2, respectively. c) and d) show a lineout at the center of the longitudinal electric field for Sim.1 and Sim.2, respectively. e) and f) represent the electron charge density for Sim.1 and Sim.2, respectively. g) and h) represent the ion charge density for Sim.1 and Sim.2, respectively. All the figures are taken after 1mm of propagation into the fully-ionized Helium plasma. The space between the dashed lines represents the area where the trapped electrons are.

Figures 3a and 3b show the envelope of the laser electric field for Sim.1 and Sim.2, respectively, after the laser has propagated 2 mm into the plasma. The front of the laser pulse is still driving non-linear plasma waves and still being self focused by this structure. The middle of the laser pulse ($1730<x_1(\mu m)<1800$) is now strongly defocussed. In Sim.1 and Sim.2, this region is where the strongest plasma waves are being driven as seen in figures 3c and 3d. Two effects determine the laser evolution here. The first is the pump depletion of the beam, and the second is the Raman side scattering (RSS). The back of the laser pulse is still in an almost plasma-free region and continues to be guided by the hollow waveguide structure seen in figures 3e-3h. However, in contrast to Sim.1, where the back of the laser pulse that resides in the hollow channel is guided but otherwise does not generate plasma waves, in the case of Sim.2, it is heavily self-modulated (figure 3b) and driving linear plasma waves (figures 3d and 3f).

Figures 4a and 4b show the transverse $k(x_2)$ spectra of the laser at $k(x_1)$ longitudinal equal to 10 (initial k of the laser – $k_0$) after propagating 1 (black), 2 (red) and 3 (blue) millimeters in the plasma for Sim.1 and Sim.2 respectively. It is clear that the laser suffers RSS since the transverse $k(x_2)$ spectrum broadens significantly as the beam propagates through the plasma [29].

Throughout the simulation box as the normalized plasma wave longitudinal electric field $eE_1/mc\omega_p$ approaches 1 (the nominal wavebreaking value for a cold plasma), copious amounts of plasma electrons are self-trapped and accelerated. Figure 5 shows the $p_1$ vs. $x_1$

phase space of all the electrons inside the simulation window for Sim.1 and Sim.2 after the laser has propagated 1 mm (5a, 5b), 2 mm (5c, 5d) and 3 mm (5e, 5f) inside the plasma. The "fingers" of charge spaced at $\lambda_p$ are indicative that the electrons are gaining energy from the plasma waves. These fingers are more prominent in Sim.2 since self-modulation due to linear plasma waves occurs for a greater portion of the laser pulse. In the case of Sim.1, the fingers are only visible in the middle of the laser pulse where the driven plasma waves self modulate the laser pulse and wavebreak to trap electrons.

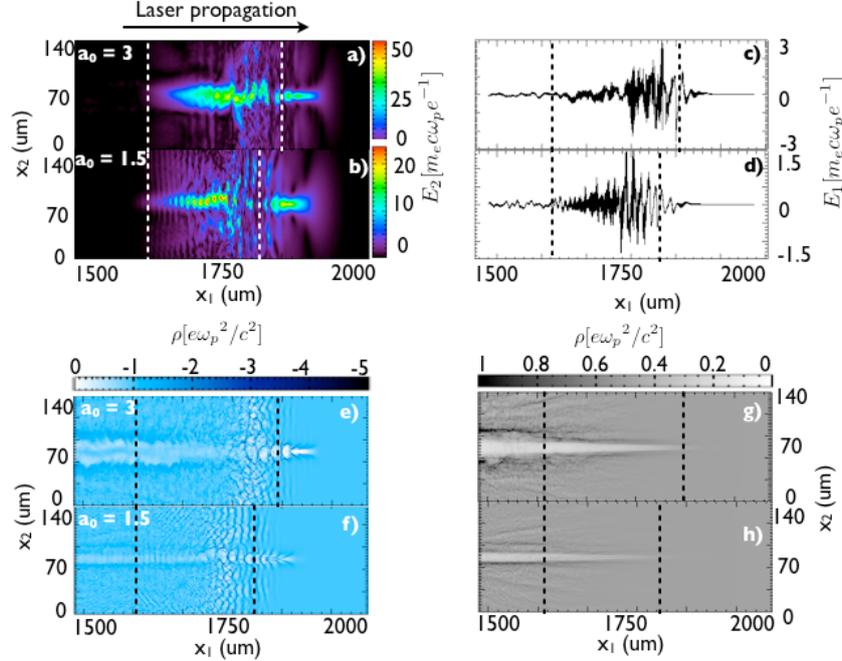

Figure 3 - a) and b) represent the laser electric field envelope for Sim.1 and Sim.2, respectively. c) and d) show a lineout at the center of the longitudinal electric field for Sim.1 and Sim.2, respectively. e) and f) represent the electron charge density for Sim.1 and Sim.2, respectively. g) and h) represent the ion charge density for Sim.1 and Sim.2, respectively. All the figures are taken after 2 mm propagation into the fully-ionized Helium plasma. The space between the dashed lines represents the area where the trapped electrons are.

At the back of the laser pulse, the plasma waves are very weak due to the electron and ion cavitation, so the electric field of the plasma waves is not strong enough to self-trap the plasma electrons. In this region, the electrons also gain energy, but the energy gain is mainly due to DLA as we will show next.

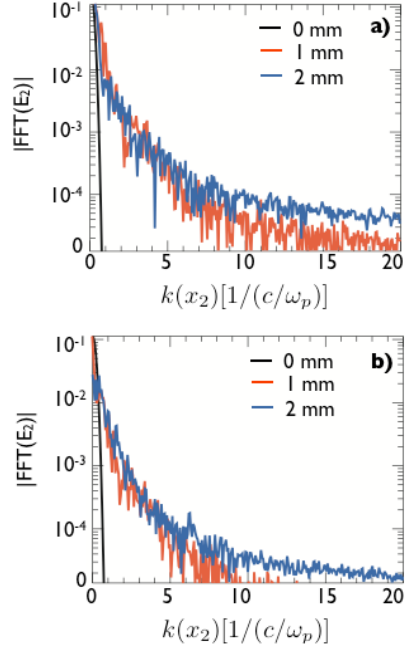

Figure 4 - a) and b) show the transverse $k(x_2)$ spectra of the laser at longitudinal $k(x_1)$ equal to 10 (initial k of the laser – $k_0$) for Sim.1 and Sim.2, respectively. The black line represents the spectrum after the laser beam propagates 0 mm into the plasma, the red 1 mm, and the blue 2 mm.

In order to investigate how electrons gain energy, we have selected a few electrons in specific locations within the plasma after running the two simulations as described in Sec. 2. We subsequently ran the same simulations again and tracked the trajectories of the selected electrons. Tracking these trajectories records the momentum, position, and fields for each simulation time step. Figures 6a-6c show the electron charge density for Sim.1 after 3 mm of propagation into the fully-ionized Helium plasma. The black dots represent the tracked electrons throughout the simulation. Here, we have selected three groups of electrons. The first group is comprised of 20 random electrons trapped (figure 6a) in the strong nonlinear plasma wave region. Their trajectories through the simulation are plotted in figure 6d where the color represents the instantaneous energy of the electrons. The second group is comprised of 20 randomly-selected electrons that are trapped in the weak, non-linear plasma wave region (figure 6b). The trajectories of this second group are shown in figure 6e. The third group is comprised of electrons trapped in the partially-hollow channel (figure 6c), where 105 random electrons were tracked, and their trajectories are plotted in figure 6f. These three groups of electrons were chosen because they represent the three different regions of the plasma where different mechanisms are at work in imparting energy to the electrons. By comparing the trajectories of these three groups of electrons, it is clear that the first group has smaller transverse oscillations because the nonlinear plasma waves are stronger and exert a higher focusing force on the electrons. In the second group, which is in the weak non-linear region of the wave, the electrons already have higher-amplitude transverse oscillations mainly due to the interaction with the transverse laser field. The oscillations of the third group are limited by the diameter of the hollow channel that is created by the electron and ion motion.

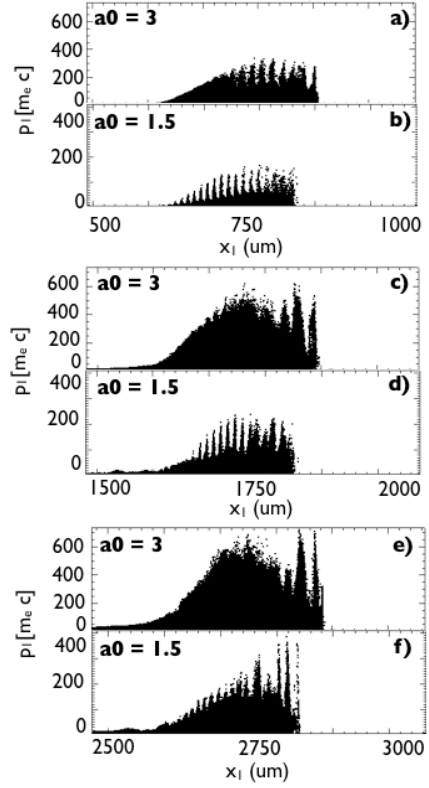

Figure 5 - a) and b) represent the phase space $x_1$ vs. $p_1$ after 1mm of propagation into the fully-ionized Helium plasma for Sim.1 and Sim.2, respectively. c) and d) represent the phase space $x_1$ vs. $p_1$ after 2 mm of propagation into the fully-ionized Helium plasma for Sim.1 and Sim.2, respectively. e) and f) represent the phase space $x_1$ vs. $p_1$ after 3 mm of propagation into the fully-ionized Helium plasma for Sim.1 and Sim.2, respectively.

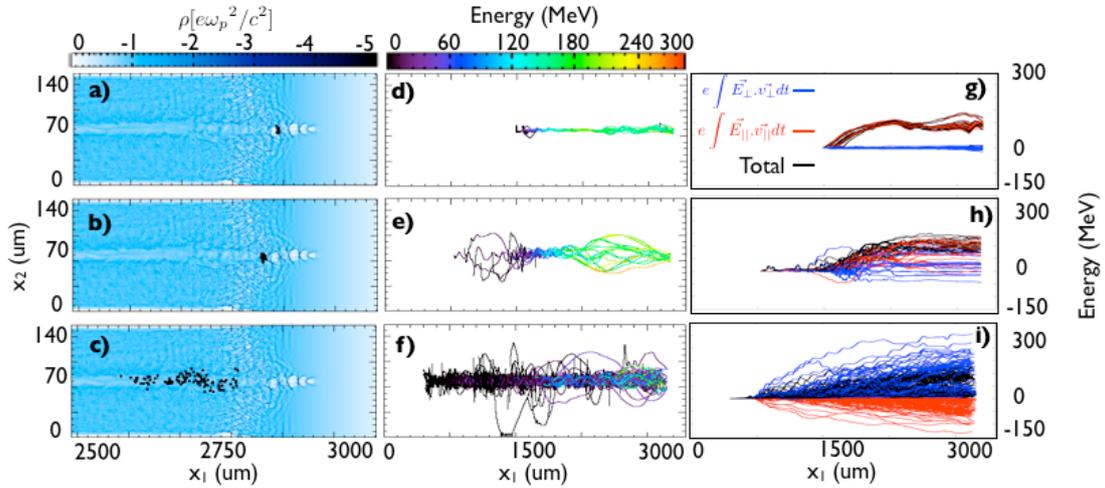

Figure 6 - a)-c) represent the electron charge density for Sim.1 after 3 mm of propagation into the fully-ionized Helium plasma. The black dots represent the position of the different tracked particles. d)-f) show the trajectories of the tracked particles throughout the simulation, and the color represents the energy of the particle. g)-i) show in blue the energy gained by the tracked electrons throughout the simulation due to the interaction with the transverse laser field. The red curve shows the energy gained by the tracked electrons throughout the simulation due to the interaction with the longitudinal electric field. The black curve shows the total energy gained by the tracked electrons throughout the simulation.

In figures 6g-6i, we show the physical mechanisms leading to energy gain for these three groups of electrons. Once the electrons are trapped, they can gain energy from the longitudinal field of the plasma wave and from DLA, which, as explained earlier, can occur when there is a considerable overlap between the trapped electrons and the laser field. The energy gained by the electrons due to the longitudinal electric field (of the plasma wave) $\mathbf{E}_{\parallel}$ was calculated for each tracked particle with the integral $e \int_0^t \vec{v_\parallel} \cdot \vec{E_\parallel} \, dt$ for every simulation time step. The energy gain due to the transverse electric field (DLA) was calculated with the integral $e \int_0^t \vec{v_\perp} \cdot \vec{E_\perp} \, dt$, where $\vec{v_\perp}$ is the velocity of the tracked electron in the plane of the laser polarization and $\vec{E_\perp}$ is the transverse field of the laser. The energy gain of the first group of electrons is presented in figure 6g, where the red lines represent the energy gain due to the longitudinal electric field, the blue lines represent the energy gain due to the transverse electric field, and the black lines represent the total energy of the electron. In this case, the dominant energy gain mechanism is the longitudinal electric field. Here, in the heavily-modulated plasma region, the laser pulse breaks up into a series of micro pulses, and each micro pulse resides in the region of the plasma wave that is decelerating for electrons, so the accelerating electrons in the plasma wave do not interact with the laser electric field. The maximum energy of the tracked electrons is 200 MeV. However, for the second group of electrons, the laser occupies a large fraction of the plasma period and interacts with the electrons during the acceleration process. Thus, in this case, the electrons gain and lose energy through both DLA and the plasma wave. Here, the tracked electrons also reach a maximum energy of 200 MeV, and the dominant acceleration mechanism is still the longitudinal electric field. The energy gain of the third group of electrons that are in the channel is completely dominated by DLA. In this case, the tracked electrons even give energy back to the longitudinal electrical field and reach a maximum energy of 190 MeV. To summarize, while the maximum energy of the electrons in all three regions is almost the same, 200 MeV, there is a transition from energy gain due to the longitudinal field of the plasma wave at the front of the drive laser to DLA towards the back of the laser pulse.

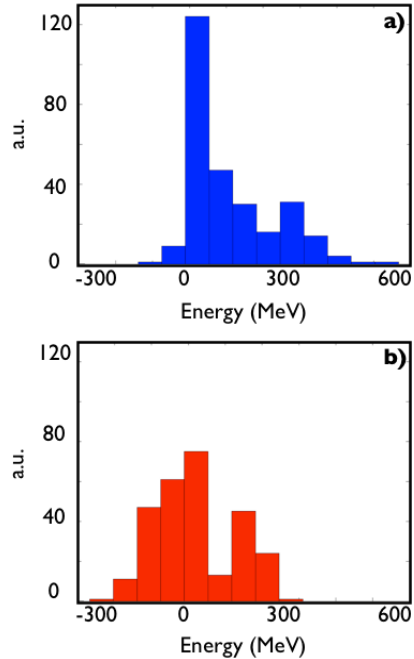

Figure 7 – Analysis of Sim.1 a) histogram that quantifies the number of particles that gain or lose energy from the transverse electrical field. b) histogram that quantifies the number of particles that gain or lose energy from the longitudinal electrical field

In figure 7, we randomly tracked 250 electrons from Sim.1 that were representative of the electron energy spectrum of the total number of particles in the simulation, and analyzed the relative efficacy of the two energy gain mechanisms for these electrons. Figure 7a shows the histogram that quantifies the number of particles that gain or lose energy from the transverse electric field. It is clear that most of the particles have a positive contribution from the transverse electric field to their final energy. The contribution of the longitudinal field is represented in figure 7b. Here, approximately 50% of the electrons lose energy and

50% gain energy from the longitudinal field. We can conclude that the transverse electric field dominates the electron energy gain for the bulk of the particles.

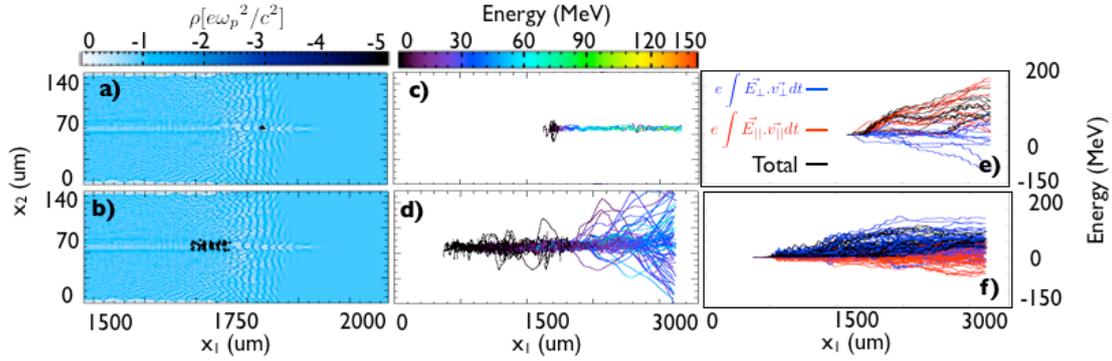

Figure 8 - a)-b) represents the electron charge density for Sim.2 after 2mm of propagation into the fully ionized Helium plasma. The black dots represent the different position of the tracked particles. c)-d) shows the trajectory of the tracked particles throughout the simulation and the color represents the energy the energy of the particle. e)-f) shows in blue the energy gained by the tracked electrons throughout the simulation due to the interaction with the transverse electric field. The red curve shows the energy gained by the tracked electrons throughout the simulation due to the interaction with the longitudinal electric field. The black curve shows the total energy gained by the tracked electrons throughout the simulation.

We repeat the analysis for Sim.2. In figure 8, we show the electron charge density after 2 mm of propagation into the fully-ionized Helium plasma. The black dots again represent the tracked electrons. Here, we have selected two groups of electrons. The first group is comprised of 10 randomly-selected electrons that are trapped in the strong nonlinear plasma wave region (figure 8a). Their trajectories throughout the simulation are plotted in figure 8c where the color represents the energy of the electrons. The second group is comprised of 66 electrons trapped in a weak non-linear plasma wave and in the partially-hollow channel region (figure 8b). The associated trajectories are shown in figure 8d. These two groups of electrons were chosen because they represent the two distinct regions within the laser pulse where different physical mechanisms are at work. By comparing the trajectories of these two groups of electrons, it is clear that the ones that are trapped in the front have smaller transverse oscillations because the nonlinear plasma waves are stronger and lead to higher focusing forces acting on the electrons (as was the case in Sim.1). The second group, located in a region where there are linear plasma waves and a small percentage of electron and ion cavitation, has higher-amplitude transverse oscillations. The amplitude of these oscillations is approximately the cavity diameter. After 2.5 mm of propagation, the transverse oscillation amplitude increases because the ion electron cavity becomes weaker and larger since the laser spot is getting larger.

The energy gain of the first group of electrons is presented in figure 8e. Here, the dominant energy gain mechanism is the longitudinal electric field. Again, the heavily-modulated laser pulse breaks up into a series of micro pulses, and each micro pulse resides in the region of the plasma wave that is decelerating for electrons, and the wake accelerated electrons do not interact with the laser electric field. In general, the transverse electric field has a negative contribution to the energy gain of the electrons because the resonance condition is not satisfied. The maximum energy of the tracked electrons in this region is 150 MeV. For the second group of electrons, the laser occupies the entire plasma period and therefore interacts with the electrons. The electrons gain and lose energy through both DLA and the plasma wave (figure 8f). However, the dominant acceleration mechanism is the transverse electric field, and the tracked electrons reach a maximum energy of 130 MeV.

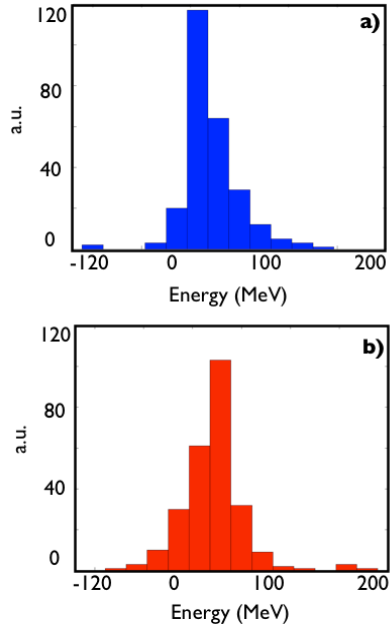

Figure 9 – Analysis of Sim.2. a) Histogram that quantifies the number of particles that gain or lose energy from the transverse electrical field. b) histogram that quantifies the number of particles that gain or lose energy from the longitudinal electrical field.

In figure 9, we tracked 250 random electrons that represented the electron energy spectrum of the total number of particles in Sim.2 and determined the dominant energy gain mechanism for the electrons. Figure 9a shows the histogram that quantifies the number of particles that gain or lose energy from the transverse electric field. It is clear that most of the particles have a positive contribution to their final energy from the transverse electric field. The contribution of the longitudinal field is represented in figure 9b. Again, we find that most of the particles have a positive contribution to their final energy from the longitudinal electric field. We can conclude that at the lower $a_0$, both mechanisms contribute roughly equally to the energy gain of the electrons. In the next section, we will analyze how these differences affect the radiation produced by the accelerated electrons.

## 5. Radiation generated in the SM-LWFA regime

In order to calculate the radiation emitted by every tracked electron, we used the post-processing code JRad [31]. This code uses the information about every tracked particle, i.e., particle trajectory, position, and momentum, to calculate the power radiated by the particle. The output is a spatially-resolved diagnostic of the energy and spectrum deposited in a virtual detector. In this JRad simulation, the detector was 13.4 mm from the end of the PIC simulation. The spatial grid of the virtual detector was 336 x 6720 μm with 100 x 2500 cells for the energy diagnostic. For the spectral diagnostic, the grid had 500 x 500 cells and 6720 μm x 117 keV for Sim.1 and 6720 um x 11.7 keV for Sim.2. Due to computational limitations, we tracked 250 random electrons that reproduce approximately the energy spectrum of all the accelerated electrons present in the simulation.

Figures 10a and 10b show the power radiated by the tracked electrons in Sim.1 and Sim.2, respectively. The FWHM divergence in the $x_2$ direction is 62 mrad for Sim.1 and 77 mrad for Sim.2. In the $x_3$ direction the FWHM divergence is 1.9 mrad for Sim.1 and 3.8 mrad for Sim.2. The divergence is larger in the $x_2$ direction because most of the electrons interact with the laser pulse, which increases their transverse excursion in the polarization plane of the laser.

The energy spectrum at $x_3 = 168$ μm for Sim.1 and Sim.2 is shown in figures 11a and 11b, respectively. For Sim.1, the spectrum peaks at 5 keV and extends ups to 100 keV where 20% of the energy is above 50 keV. For Sim.2, the spectrum peaks at 0.2 keV and extends up to 12 keV. Figures 11c and 11d show in red a lineout of the radiation spectrum at $x_2 = 3360$ µm. The black dashed curves show a fit to the synchrotron asymptotic limit (SAL) function [44] $(E/E_c)^2 K^2_{2/3}(E/E_c)$, for radiation emitted along the axis, where $K^2_{2/3}$ is the modified Bessel function, E is the energy of the emitted radiation, and $E_c$ is the critical energy that is the value at which roughly 50% of all the radiated energy is contained. We tried fitting the SAL function to the red curve in figures 11c and 11d, but the fit was not adequate. We then fitted the sum of two SAL functions, one that fitted the peak and another that fitted the high-energy tail of the energy spectrum. For Sim.1, the best fit gave an $E_c = 13$ keV for the lower energy part of the spectrum and an $E_c = 45$ keV for the high-energy tail (figure 11c). For Sim.2, the best fit for the sum of two SAL functions gave an $E_c = 0.3$ keV and an $E_c = 1.4$ keV (figure 11d). The fact that we had to use the sum of two SAL functions to properly fit the data implies that there must be two distinct groups of electrons that give rise to two different synchrotron spectrums- one for the main body of the electrons and the other for the high-energy tail. Possibly, the electrons that are dominated by DLA are responsible for the high-energy tail of the spectrum since they have the similar energies to the electrons dominated by the wakefield, but have larger oscillation radius, thus radiating higher energy photons. The lower-energy part of the energy spectrum, in turn, is possibly dominated by electrons accelerated by a combination of both the longitudinal field and DLA. This hypothesis requires further careful examination.

It is noteworthy that even though the SAL function relies on a simplistic assumption that the radiating electrons have a constant energy and radius of oscillation, it is rather remarkable that a simple sum of two SAL functions can reproduce the simulated radiation energy spectrums.

The almost ten times difference of the maximum radiated energy between Sim.1 and Sim.2 can be explained by the expression for energy loss of electrons to radiation, $W_{loss} \propto n_e^2 \gamma^2 r_\beta^2$ [44] where $\gamma$ is the energy of the electron and $r_\beta$ is the radius of oscillation of the electron. Since the plasma density in these two simulations is the same, the two factors that will influence the radiated energy spectra are energy $\gamma$ and the distance from the axis $r_\beta$. In Sim.1, the tracked electrons used to calculate the radiated energy spectrum have energies up to 280 MeV and in Sim.2 energies up to 110 MeV. This difference in the maximum energy would already give a six-times difference in the energy loss in the two cases. Also, the average radius of oscillation of the tracked electrons for Sim.1 is 1.5 times larger than in Sim.2. Taking these two effects into account, we would expect a nine times difference in the maximum energy lost to radiation between Sim.1 and Sim.2. This estimate is close to the approximate ten times difference of the maximum radiated energy observed between figures 11a and 11b.

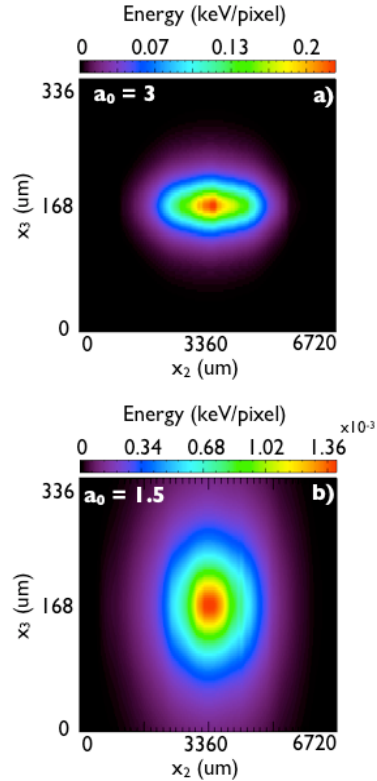

Figure 10 - a) and b) shows the power radiated of the tracked electrons in Sim.1 and Sim.2, respectively through a perpendicular plane detector. The color scale represents the energy (keV/pixel) deposited in each cell.

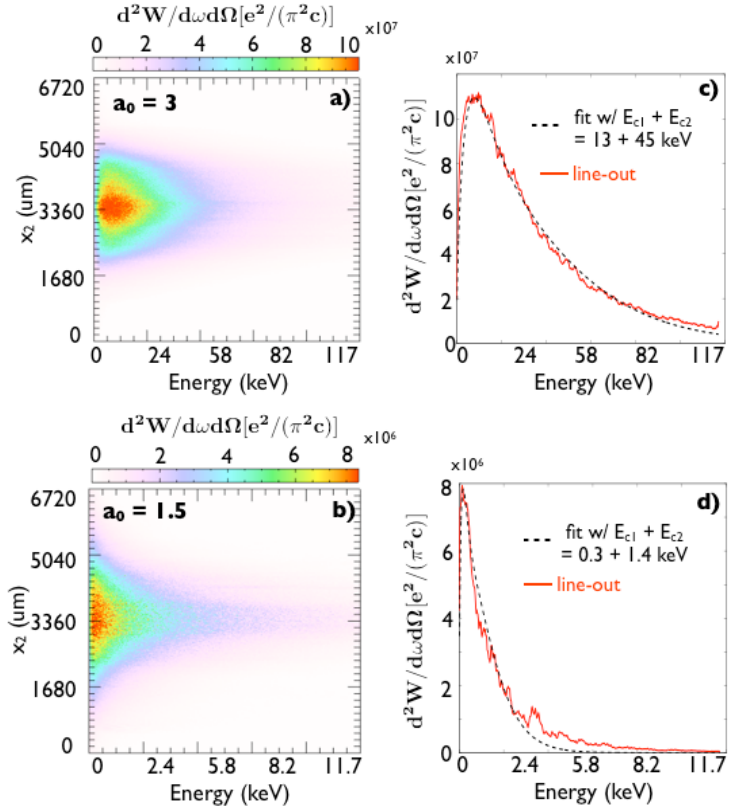

Figure 11- a) and b) show the energy spectrum of the tracked electrons in Sim.1 and Sim.2, respectively, at $x_3$ = 168 um. The color scale represents the energy/frequency/solid angle deposited in each cell. c) and d) show in red a lineout of a) and b) at $x_2 = 3360$ μm, respectively. The black dashed curves show a fit to the function $(E/E_{c1})^2 K^2_{2/3}(E/E_{c1}) + (E/E_{c2})^2 K^2_{2/3}(E/E_{c2})$, where $E_{c1}$ = 13 keV and $E_{c2}$ = 45 keV in c) and $E_{c1}$ = 0.3 keV and $E_{c2}$ = 1.4 keV in d).

## 6. Conclusions

In conclusion, we have explored through 2D PIC simulations and the JRAD code the possibility of using self-modulated laser wakefield accelerator as an x-rays source. We have presented results with moderate values of $a_0$ (1.5 and 3) that are likely to be achieved in the ps, high-energy lasers being constructed in the major laboratories where HEDS work is being carried out. Because of the long laser pulse lengths and the long propagation lengths through the plasma, we have used a 2D PIC code but with mobile ions. In both case considered, the plasma ions move in the transverse direction in the back of the pulse and form a partially-evacuated channel. In the front of the pulse, strong laser self modulation leads to trapping and acceleration of electrons by the longitudinal field in both cases. In the back of the pulse, DLA plays a dominant role for the higher $a_0$ case (Sim.1) whereas both the longitudinal field of a quasi-linear plasma wave and DLA contribute to electron heating in the low $a_0$ case (Sim.2). In both cases, strong Raman side-scattering is observed in the middle of the pulse. Electrons were accelerated in the forward direction up to an energy of 300 MeV in Sim.1 and produced forward-directed x-ray radiation up to 100 keV with a relatively small FWHM divergence of 62 x 1.9 mrad. Electrons were accelerated up to an energy of 190 MeV in Sim.2 and produced forward directed x-ray radiation up to 12 keV with a FWHM divergence of 77 x 3.8 mrad. The radiation energy spectrums of Sim.1 and Sim.2 can be well fitted with a sum of two SAL functions suggesting that these arise from radiation emitted by two groups of electrons that comprise the overall accelerated spectrum.

**Acknowledgments**

Work supported by DOE grant DE-SC0010064. Simulation work done on the Hoffman2 Cluster at UCLA and on NERSC. The work of J. L. Martins was financially supported by the European Research Council (ERC − 2010 − AdG Grant 267841). The authors also wish to acknowledge the computing facilities where the post-processing were done: the SuperMUC supercomputer (through PRACE) at the Leibniz Supercomputing Centre in Germany and the cluster ACCELERATES in Instituto Superior Tecnico in Lisbon, Portugal.

**References**

[1] Basic research needs for high energy density laboratory physics, Report of the workshop on high energy density laboratory physics research needs, November 15-18 2009
[2] Next Generation Photon Sources for Grand Challenges in Science and Energy, A report of a sub committee to the BES advisory committee, May 2009
[3] Hurricane O A, et al. 2014 *Nature* **506** 343–8
[4] Hicks D, et al. 2012 *Phys. Plasmas* **19** 122702
[5] Benuzzi-Mounaix A, et al. 2011 *Phys. Rev. Lett.* **107** 165006
[6] Kritcher A L, et al. 2008 *Science* **322** 69
[7] Glenzer S H and Redmer R, 2009 *Rev. Mod. Phys.* **81** 1625
[8] Pak A, et al. 2004 *Rev. Sci. Instrum.* **75** 3747
[9] Wang S, Clayton C E, Blue B E, Dodd E S, Marsh K A, Mori W B and Joshi C 2002 *Phys. Rev. Lett.* **88** 135004
[10] Wang X, et al. 2009 *Phys. Rev. Spec. Top.-Ac.* **12** 051303
[11] Johnson D K, et al. 2006 *Phys. Rev. Lett.* **97** 175003
[12] Albert F, et al. 2013 *Phys. Rev. Lett.* **111** 235004
[13] Rousse A, et al. 2004 *Phys. Rev. Lett.* **93** 135005
[14] Corde S, Ta Phuoc K, Lambert G, Fitour R, Malka V and Rousse A, 2013 *Rev. Mod. Phys.* **85** 245
[15] Kneip S, et al. 2010 *Nat. Phys.* **6** 980
[16] Kneip S, et al. 2008 Observation of Synchrotron Radiation from Electrons Accelerated in a Petawatt-Laser-Generated Plasma Cavity *Phys. Rev. Lett.* **100** 105006
[17] Esarey E, Krall J and Sprangle P 1994 *Phys. Rev. Lett.* **72** 2887
[18] Andreev N E, Gorbunov L M, Kirsanov V I, Pog-osova A A and Ramazashvili R R 1992 *Pisma Zh.Eksp.Teor.Fiz.* **55**, 551
[19] Antonsen T M Jr. and Mora P 1992 *Phys. Rev. Lett.* **69** 2204
[20] Sprangle P, Esarey E, Krall J and Joyce G 1992 *Phys. Rev. Lett.* **69** 2200
[21] Joshi C 1981 *Phys. Rev. Lett.* **47** 1285-8
[22] Modena A, et al. 1995 *Nature* **377** 606–8
[23] Pukhov A, Sheng Z M and Meyer-ter-Vehn J 1999 *Phys. Plasmas* **6** 2847
[24] Shaw J L, Tsung F S, Vafaei-Najafabadi N, Marsh K A, Lemos N, Mori W B and Joshi C 2014 *Plasma Phys. Contr. F.* **56** 084006
[25] Gordon D, et al. 1998 *Phys. Rev. Lett.* **80** 2133-6
[26] Zhang X, Khudik V N, and Shvets Gennady 2015 *Phys. Rev. Lett.* **114** 184801
[27] Albert F, to be submitted to *Phys. Rev. Lett*

[28] Fonseca R A, et al. 2002 *Lect. Notes Comput. Sc.* (ICCS 2002, LNCS 2331, edited by Sloot P M A et al.) p. 342-51
[29] Tzeng, et al. 1996 *Phys. Rev. Lett.* **78** 3332
[30] Mangles S P D, et al. 2005 *Phys. Rev. Lett.* **94** 245001
[31] Martins J L, Martins S F, Fonseca R A and Silva L O 2009 Harnessing Relativistic Plasma Waves as Novel Radiation Sources from Terahertz to X-Rays and Beyond *SPIE Conference* **7359** 73590V
[32] Sprangle P, Esarey E and Ting A 1990 Nonlinear Theory of Intense Laser-Plasma Interactions *Phys. Rev. Lett.* **64** 2011–4
[33] Sun G Z, Ott E, Lee Y C and Guzdar P 1987 *Phys. Fluids* **30** 526
[34] Coverdale C A, et al. 1995 *Phys. Rev. Lett.* **74** 4659
[35] Sprangle P, Tang C M and Esarey E 1987 *IEEE T.Plasma Sci.* **15** 145
[36] Wilks S C, et al. 1989 *Phys. Rev. Lett.* **62** 2600
[37] Shaw J, "Satisfying the direct laser acceleration resonance condition in a laser wakefield accelerator" to be published in the proceedings of the AAC 2014
[38] Palmer R B 1972 *J. Appl. Phys.* **43** 3014
[39] Courant E, Pellegrini C and Zakowicz W 1985 *Phys. Rev. A* **32** 2813
[40] Whittum D, Sessler A and Dawson J 1990 *Phys. Rev. Lett.* **64** 2511
[41] Jackson J D, Classical Electrodynamics
[42] Joshi C, Clayton C. E., and Chen F. F. 1982 Phys. Rev. Lett. **48**-13, 874-877
[43] Mori W B, et al. 1994 *Phys Rev. Lett.* **73** 1482
[44] Esarey E, Shadwick B A, Catravas P and Leemans W P 2002 *Phys. Rev. E* **65** 056505